\documentclass{elsart}
\usepackage{amssymb}
\usepackage{epsfig}
\psfigdriver{dvips}

\begin{document}

\begin{frontmatter}

\title{Investigation of vortex dynamics in Josephson junction arrays with magnetic flux noise measurements}%
\author{S. Candia, Ch. Leemann, S. Mouaziz, P. Martinoli}
\address{Institut de Physique, Universit\'{e} de Neuch\^{a}tel, 2000 Neuch\^{a}tel, Switzerland}

\begin{abstract}
We have measured the magnetic flux noise in a two-dimensional
Josephson junction array in nominally zero magnetic field, in the
vicinity of the superconducting transition. This transition is of
the Berezinski-Kosterlitz-Thouless (BKT) type, driven by the
unbinding of thermally excited vortex-antivortex pairs. At
temperatures just above the BKT  transition temperature $T_{c}$,
the flux-noise power spectrum $S(\omega)$ is white up to a
temperature dependent cross-over frequency $\omega_{\xi}(T)$. We
identify $\omega_{\xi}$ with the characteristic time related to
the scale at which the cross-over from free vortex response to a
response dominated by bound pairs occurs. The signature of free
vortices is thus white noise while bound vortex pairs give rise to
a $1/\omega$ dependence.
\end{abstract}

\end{frontmatter}
\section{Introduction}

The Berezinski-Kosterlitz-Thouless (BKT) transition \cite{BKT} is
the subject of many investigations, both experimental and
theoretical \cite{mooij,ahns,minn,chris}. Two-dimensional
Josephson junction arrays (JJA) in zero magnetic field represent
an excellent experimental system to investigate the BKT
transition. The easily controllable parameters in the fabrication
process, and the precision achieved with photolithographic
techniques make good quality and reproducible samples available.

Many experimental techniques are used to investigate the nature of
the transition \cite{newrock}. Most of them consist of measuring
the sample by means of invasive methods, and to take the limit of
zero drive excitation. In this work we use a non invasive method,
which provides a 'clean' way of measuring the transition.

By measuring the magnetic flux noise in a triangular JJA we have
obtained the frequency power spectrum $S_{\phi}(\omega)$ at
various temperatures in the transition zone. We relate the
different characteristics of $S_{\phi}(\omega)$ with the principal
features of the transition.

\section{Theoretical background}

Two-dimensional JJA constitute physical realisations of the
XY-model, they are therefore ideally suited for an experimental
investigation of the BKT transition. The analysis of experimental
data is only slightly complicated by the fact that the single
junction coupling energy $E_{J}(T)$ is temperature dependent,
necessitating the introduction of a reduced temperature
$\tau=k_{B}T/E_{J}(T)$, where $k_{B}$ is Boltzmann's constant.

In zero magnetic field, the ground state of a JJA at zero
temperature is completely ordered, i.e. all the phases are aligned
\cite{trans}. At any non-zero temperature, thermal fluctuations
give rise to different kinds of excitations which destroy long
range phase coherence. The first excitations to appear are spin
waves. When the temperature fluctuations become more important,
vortex-antivortex (V-AV) pairs are created in the system. The
array remains superconducting (existence of quasi long range
order) up to the transition temperature $T_{c}$. At $T_{c}$ the
largest pairs of V-AV dissociate, and the first free vortices
appear. Above $T_{c}$ the free vortex density will raise with
increasing temperature and is proportional to $\xi_{+}^{-2}$,
where $\xi_{+}(\tau)\propto\exp[b/(\tau-\tau_{c})^{1/2}]$ is the
BKT correlation length. In this regime the response of the system
at sufficiently long length scales [compared with $\xi_{+}(T)$] is
determined by free vortices, while V-AV pairs are responsible of
the behavior at shorter length scales. For a simple diffusion
process $ r_{\omega} \approx (D/\omega)^{1/2}$, where D is the
diffusion constant, is the typical diffusion distance in time
$\omega^{-1}$\cite{ahns,at}. Then, by measuring the system for a
given time scale (\textit{i.e.}, at a given frequency $\omega$) we
are testing a spatial scale $r_{\omega}$. In the frequency domain,
the low frequency response comes from the contribution of signals
extending over large scales, while higher frequencies are
associated with smaller distances. At $T=T_{c}$ the correlation
length $\xi_{+}(T)$ approaches infinity, which in the frequency
domain corresponds to the limit $\omega=0$. Above $T_{c}$ pairs of
characteristic size $\xi_{+}(T)$ dissociate. We can therefore
expect that the frequency $\omega_{\xi}$ associated with the
length scale $\xi_{+}(T)$ will mark a regime crossover. Finally,
we notice the exponentially decaying behavior of $\xi_{+}(T)$,
implying that $\omega_{\xi}(T)$ should be an increasing function
of T.

The flux noise power spectrum $S_{\phi}(\omega)$ (\textit{i.e.},
the Fourier transform of the flux-flux correlation function)
permits us to characterize the behavior of the vortex fluctuations
over a very broad frequency range. Many theoretical works can be
found in the literature which predict different dependencies for
$S_{\phi}(\omega)$. All of these models predict two regimes: at
low frequencies a frequency independent white noise which becomes
$\omega^{-\alpha} (\alpha > 0)$ dependent at higher frequencies.
Citing only some of these works, the predicted value for $\alpha$
varies among $3/2$, $1$ or $2$, depending on the assumed model
\cite{hwang,tiesinga,wagen,houlrik}.

\section{Results and discussion}

The power spectra of a proximity effect Pb/Cu/Pb triangular JJA at
a number of temperatures close to $T_{c}$ were measured. With
physical vapor deposition and subsequent photolithography
star-shaped islands of Pb were deposited  on a Cu layer of $\sim
2000 \AA$ \cite{chris}. The lattice constant $a$ is $15 \mu m$,
and the junction width is $1\mu m$.

\begin{figure}[htb]
\centering
\includegraphics[angle=0,width=100mm]{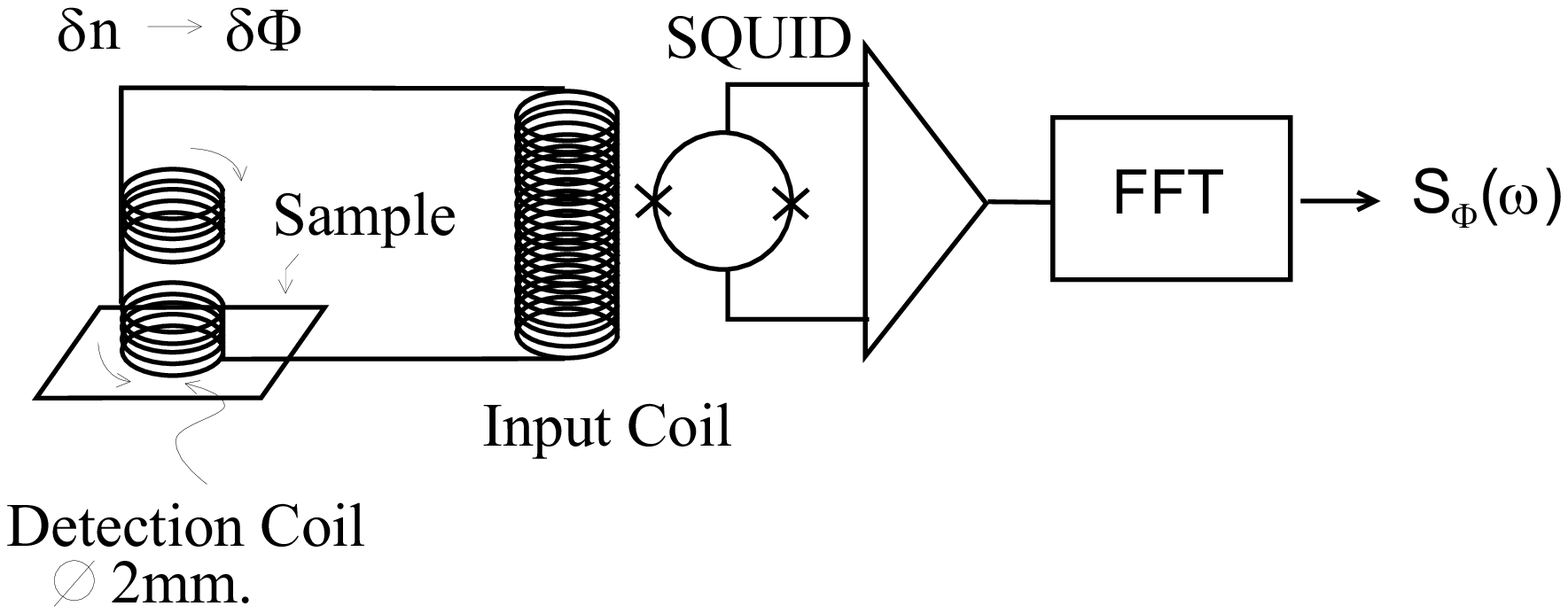}
\caption{Experimental configuration: an astatic coil is mounted
over the sample to detect magnetic flux noise fluctuations. A DC
SQUID amplifies the fluctuations which are then fed into the FFT.}
\label{fig1}
\end{figure}

The experimental setup is shown in Fig. \ref{fig1}. An astatic
detection coil was mounted over the sample, the first winding
lying $\sim 20 \mu m$ above the sample. A DC-SQUID detects
magnetic flux fluctuations seen by the coil, and transforms them
in voltage. A dynamic signal analyzer collects and analyzes the
SQUID response, and makes a fast Fourier transform (FFT) giving
directly the power spectrum $S_{\phi}(\omega)$ of the signal.

\begin{figure}[htb]
\centering
\includegraphics[angle=0,width=100mm]{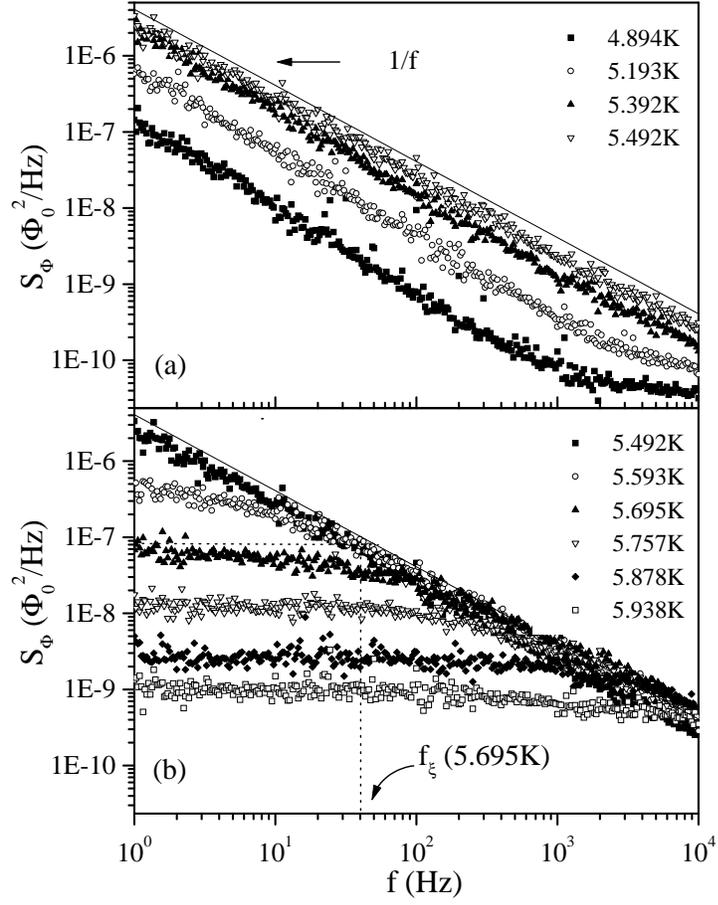}
\caption{Flux noise as a function of frequency f at temperatures
slightly below (a) and slightly above (b) the BKT transition. On a
log-log plot straight lines have slope -1.}
\label{fig2}
\end{figure}

The resulting power spectra as a function of frequency
$f=\omega/2\pi$ at different temperatures below the BKT transition
temperature are presented in Fig. \ref{fig2}a, a selection of
spectra above $T_{c}$ is shown in Fig. \ref{fig2}b. Above $T_{c}$,
for frequencies $\omega$ up to a characteristic temperature
dependent frequency $\omega_{\xi}$ the noise is Johnson like (or
white). For $\omega>\omega_{\xi}$ the noise becomes proportional
to $1/\omega$. If we consider free vortices as uncorrelated
entities, the white noise spectrum observed for
$\omega<\omega_{\xi}$ appears to be a reasonable result. At
smaller length scales the dynamics is strongly influenced by the
presence of V-AV pairs, and a different type of spectrum is
expected. As evidenced by our results this spectrum is of the
$1/\omega$ form. It should be emphasized that we find both
$\omega^{0}$ and $\omega^{-1}$ spectra, while other like
$\omega^{-2}$ or $\omega^{-3/2}$ are not observed.

In the following we focus on the free vortex dominated white noise
contribution to $S_{\phi}(\omega)$ below $\omega_{\xi}$ (the
behavior of $S_{\phi}(\omega)$ associated with V-AV pairs will be
discussed elsewhere). From the fluctuation-dissipation theorem
relating $S_{\phi}(\omega)$ to the real part of the conductivity,
it is straightforward to show, using a Drude-like approach to
describe the free vortex 'charges', that at low frequencies

\begin{equation}
S_{\phi}(\omega) = CT/n_{v}(T)
\label{noise}
\end{equation}

where $n_{v}(T)$ is the free vortex areal density and C a
temperature independent constant. In a very narrow critical
temperature range above $T_{c}$ one expects $n_{v}(T) \propto
\xi_{+}^{-2}(T)$, whereas a simple activated behavior $n_{v}(T)
\propto \exp(-\Delta/kT) = \exp(-B/\tau)$ should become manifest
outside the critical region, at temperatures well above $T_{c}$.
The second exponential form relies on the observation that the
energy $\Delta$ required to nucleate a vortex is proportional to
$E_{J}(T)$. In Fig. 3 the quantity $T/S_{\phi}(\omega)$ extracted
from the white noise 'plateaus' of Fig. \ref{fig2}b is plotted
logarithmically as a function of $1/\tau$. The data exhibit an
almost Arrhenius-like behavior with a vortex nucleation energy
$\Delta = BE_{J} \approx 1.6E_{J}$. There is no trace of critical
behavior [$n_{v}(T) \propto \xi_{+}^{-2}(T)$], which should be
reflected in a downward curvature of the data as $\tau$ approaches
$\tau_{c}$. This indicates that, in order to observe genuine
critical behavior, $S_{\phi}(\omega)$ should be measured at
frequencies much lower than those accessible in our experiments.
This observation rises some questions regarding the validity of
the analysis, based on $r_{\omega}\approx \xi_{+}(T)$ to study
$\omega_{\xi}$, carried out in \cite{shaw}.

\section{Conclusions}

From the results presented here we conclude that it is possible to
characterize the features of the BKT transition by inspecting the
flux noise spectra for different temperatures. When $T>T_{c}$ the
spectrum presents two easily distinguishable parts: for
frequencies $\omega$ lower than a crossover frequency
$\omega_{\xi}(T)$ we observe white noise, the signature of free
vortices, while for $\omega\geq \omega_{\xi}(T)$ the spectrum
decays as $1/\omega$, signature of bound V-AV pairs. As the
temperature grows, the characteristic frequency $\omega_{\xi}(T)$
increases, as expected. At $T_{c}$ an infinite vortex correlation
length implies that the crossover frequency vanishes and the
$1/\omega$ spectrum is present at all frequencies. The same
functional frequency dependence is observed for temperatures
slightly below $T_{c}$, where the intensity of the noise decreases
with T.

\section{Acknowledgments}

We thank S.E. Korshunov for useful discussions and critical
reading of this paper. This work was supported by the Swiss
National Science Foundation, the Swiss Federal Office for
Education and Science within the framework of the TMR network
Superconducting Nanocircuits of the European Union and the Vortex
Program of the European Science Foundation.

\pagebreak
\newpage

\pagebreak
\newpage

\pagebreak
\newpage

\begin{figure}[htb]
\centering
\includegraphics[angle=-90,width=100mm]{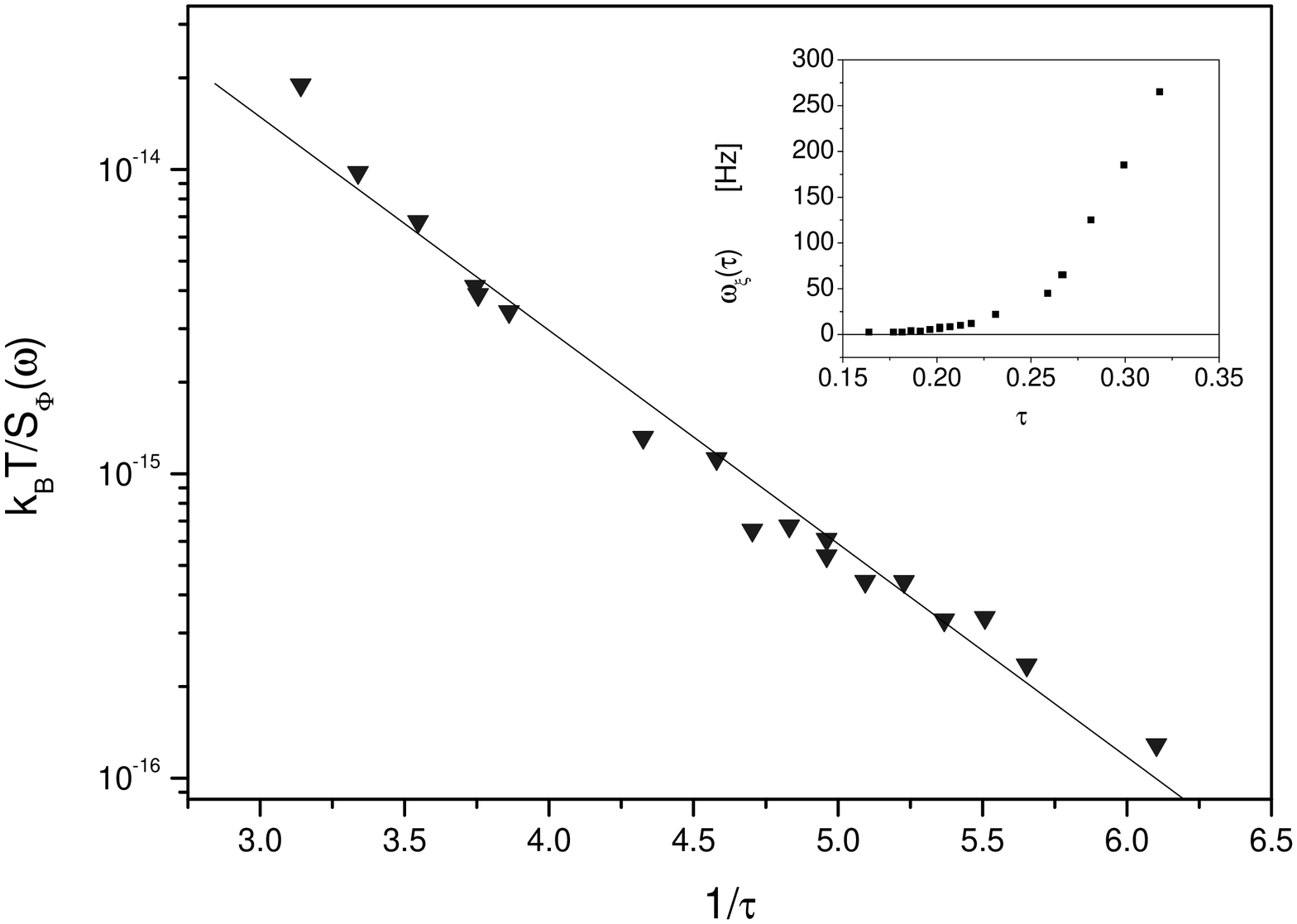}
\caption{Arrhenius plot of $k_{B}T/S_{\phi}(\omega)$ as deduced
from the noise 'plateaus' of Fig. \ref{fig2}b at
$\omega<\omega_{\xi}$. The straight line is a fit through the
data. The temperature dependence of the crossover frequency
$\omega_{\xi}$ is shown in the inset.}
\label{fig3}
\end{figure}

\end{document}